# Factors that affect the technological transition of firms toward the industry 4.0 technologies[1]

**Seung Hwan Kim**[a], **Jeong hwan Jeon**[b,c,g], **Anwar Aridi**[d], **Bogang Jun**[e,f,g]

[a]Technology Management, Economics and Policy Program, Seoul National University, Seoul, South Korea
[b]Department of Electrical Engineering, Ulsan National Institute of Science and Technology, Ulsan, South Korea
[c]Graduate School of Artificial Intelligence, Ulsan National Institute of Science and Technology, Ulsan, South Korea
[d]Finance, Competitiveness, and Innovation Global Practice, Europe and Central Asia Unit, The World Bank Group
[e]Department of Economics, Inha University, South Korea
[f]Department of Data Science, Inha University, South Korea
[g]Research Center for Small Businesses Ecosystem, Inha University, South Korea

**Abstract**

This research aims to identify factors that affect the technological transition of firms toward industry 4.0 technologies (I4Ts) focusing on firm capabilities and policy impact using relatedness and complexity measures. For the analysis, a unique dataset of Korean manufacturing firms' patent and their financial and market information was used. Following the *Principle of Relatedness*, which is a recently shaped empirical principle in the field of economic complexity, economic geography, and regional studies, we build a technology space and then trace each firm's footprint on the space. Using the technology space of firms, we can identify firms that successfully develop a new industry 4.0 technology and examine whether their accumulated capabilities in their previous technology domains positively affect their technological diversification and which factors play a critical role in their transition towards industry 4.0. In addition, by combining data on whether the firms received government support on R&D activities, we further analyzed the role of government policy in supporting firms' knowledge activity in new industry 4.0 technologies. We found that firms with higher related technologies and more government support are more likely to enter new I4Ts. We expect our research to inform policymakers who aim to diversify firms' technological capabilities into I4Ts.

*Keywords:* Industry 4.0, Economic Complexity, Patent Data, Knowledge Accumulation Strategy of a Firm, Relatedness

---

[1] This research is an output of the World Bank Seoul Center for Finance and Innovation

# 1. Introduction

In the last half-century, South Korea underwent a fast structural change in the economy from an agricultural to an industrialized society. According to Wade (2018), achieving industrialization is not an easy task. Only a few countries outside Europe and European offshoots (i.e., Australia, New Zealand, Canada, and the United States) have achieved industrialization by escaping the Malthusian trap. Excluding city-states such as Singapore and Hong Kong, very few countries in the world, like Japan, Taiwan, and South Korea, have achieved sustainable economic development, which was capable of transforming their economy from a backward stage to an advanced industrialized society (Gerybadze, 2018). This study focuses on the Korean case of industrialization, especially the recent experience of the Fourth Industrial Revolution.

Scholars have argued that we are living in the era of the Fourth Industrial Revolution (Liao et al., 2017; Lu, 2017), although there exists scholarly debate about the discontinuity of the Fourth Industrial Revolution (Rifkin, 2011; Nuvolari, 2019). Living in the era of the industrial revolution implies that the windows of opportunity for economic development are now open for developing countries (Perez, 1988; Perez, 2003). According to Perez (2003), the technological revolution would provide the best opportunities for catching up with technical changes sufficient for initiating and advancing the development process. This is because every country is a beginner at the early stage of a new techno-economic paradigm, and the probability of success by leap-frogging is increased in this stage.

Perez (2003) argued that each technological revolution is a cluster of technological systems. For example, during the Second Industrial Revolution around 1910, the mass production revolution and its successive systems allowed the economy to achieve structural change, while the Third Industrial Revolution around the 1970s was associated with the revolution in information technology. The recent industrial revolution is associated with industry 4.0 (I4) (Liao et al., 2017; Lu, 2017; Popkova et al., 2019)[2]. I4 technologies are expected to provide the domain knowledge that affects a wide range of the economy and, therefore, can be regarded as the core technologies in the Fourth Industrial Revolution.

---

[2] In fact, the term "industry 4.0" was introduced as part of German industrial policy that aimed to improve the production system by combining the IT system to the current manufacturing system. However, "industry 4.0" is often used interchangeably with the Fourth Industrial Revolution, since the new industries that emerged during the fourth industrial revolution are yet to be defined.

Firms were the main actor that created new technologies and new sectors during the technological revolutions. It is true that the National Innovation System (NIS) should be considered when we examine the factors that determine the success of the revolution (Freeman, 1987; Lundvall, 1992). Lundvall (1992) argued that innovation at the country level depends on their NIS, consisting of not only enterprises, but also research institutes, universities, and government. In a broader perspective, it even includes the accumulated human capital, dynamics of the labor market, learning among enterprises, and sets of policies. Kim et al. (2000) also described that human capital formation, the inflow of foreign technologies, government's various policies provide the environment for firms' innovation activities. Within this environment, firms develop their strategies at the microeconomic level to sustain growth by acquiring technological capability. In the long run, these firms' innovation activities work as the main engine of the economic development of a country by leading the emergence of new sectors (Saviotti and Pyka, 2013; Saviotti et al., 2016). In this study, therefore, we focus on the main actor in creating innovation: a firm.

What are the characteristics of I4 technologies and how do firms enter the I4 technology space? What factors support firms in their technological diversification towards the I4 technologies? Further, do technology policies support firms to leverage emerging technological opportunities to achieve economic development and make their economy more advanced in the era of the Fourth Industrial Revolution? This paper aims to identify the factors that affect the technological transition of firms toward advanced technologies, especially those associated with industry 4.0, by looking at the technological trajectories of Korean firms. The core questions are (i) what are the characteristics of I4 technologies and how these are changed in terms of complexity, (ii) among the three factors of technological relatedness, complexity, and government policy, which play a critical role in firms' technological diversification, especially when firms enter the I4 technology space, and (iii) which type of firm is more likely to succeed in entering a new technology that is associated with I4 technologies. The target audience of this research are businesses and business representatives and associations that aims to diversify their technological capabilities towards I4T, researchers, and policymakers who aim at designing and implementing innovation policies that supports firms' and industry diversification and transformation.

To explore these questions, the research adopts a methodology from the *principle of relatedness*. The principle indicates that firms, cities, regions, and countries are more likely to undertake new economic activities such as those related to new technologies, new products, and new industries when they already conduct related activities (Hidalgo et al., 2018; Hidalgo, 2021). In their seminal work, Hidalgo et al. (2007) constructed product space using world trade data and tracked the trajectories of industrial diversification of countries in the space. Following Hidalgo et al. (2007), we build the technology space of Korean firms and follow the footprints of technological diversification of firms focusing on the transition toward industry 4.0. patent data, together with data on firms' financial information and Korean government policy allow us to examine various factors that affect the technological diversification of firms.

## 2. Literature review: Technological diversification of firms

Diversification is one of the characteristics of modern enterprises (Chandler, 1990; Chandler Jr, 1993; Chandler et al., 2009). Scholars have found various factors that affect the pattern and the probability of success of diversification. For example, the firm's size (Freeman, 2013) or age (Huergo and Jaumandreu, 2004) can affect technological diversification. Further, the quantitative aspect of R&D investment, such as the total amount (Griliches, 1998), and the qualitative aspect of R&D investment, such as the persistence of the investment, also affect the firms' technological expansion. Factors associated with human capital, such as the CEO's expertise in technology and collaborative environment among R&D labors, also affect technological diversification. In addition, firms' technological strategies, for example, either exploitation or exploration, can affect the pattern of diversification. Given those factors, firms expand their technological boundaries resulting from the interaction among them rather than a stochastic matter of a single factor (Weaver, 1991; Hidalgo, 2021).

Teece (1980, 1982) explained the firms' technological diversification by building the theory of multi-product firms (i.e., firms with a diversified portfolio of related products). He argued that when a firm develops a product requiring proprietary know-how and specialized physical assets, it tends to choose an efficient way to organize its economic activities, resulting in the diversification of its activities. Given the fungible and tacit characteristics of organizational knowledge, profit-seeking firms diversify in a way that avoids the high transaction costs associated with trading services and specialized assets in various markets. The direction of the diversification, however, is not random but shows a path-dependent pattern. Analyzing US firm data from 1987,

Teece et al. (1994) showed that the most common way of diversification is by adding related activities.

Focusing on the technological diversification of firms, Jaffe (1986) introduced a measure of a firm's technological distance by examining its patents. To characterize the technological position, Jaffe (1986) used the distribution of firms' patents by patent class and defined the cosine similarity index that represents the change in distribution over time. He found evidence that firms' patents, together with their profit and market value, are systematically related to the technological position of their research programs, and that movement in the technology space follows the pattern of contemporaneous profits according to different technological positions. Breschi et al. (2003) also studied the technological diversification of a firm by introducing technological relatedness, focusing on the development of its core technology. They calculated the cosine similarity index by examining the co-occurrence of International Patent Classification (IPC) codes in every patent and found that knowledge relatedness, measured using the cosine similarity index, is a critical factor in firms' technological diversification.

However, the methodology used in previous studies (e.g., using cosine similarity and defining relatedness centering on core activities) has room to be improved. First, the process of aggregating all technologies to obtain the technological relatedness of the whole industry and calculating firms' technological relatedness based on their core technology does not reflect firms' heterogeneous characteristics. Similarly, applying industry-level technological relatedness to measure within-firm technological relatedness is rather vague, as the level cannot consider the firm's accumulated technological infrastructure or path-dependent characteristics. Moreover, defining a firm's core technology as the highest proportion of the IPC could be an artificial interpretation of the analysis. As the result combines several core technologies, it is hard to differentiate the unique core technology. Hence, it is necessary to develop a measure of relatedness involving a firm's idiosyncratically accumulated technological infrastructure.

To better capture the firms' technological trajectories based on what the technologies they already own, we use the methodology from the *principle of relatedness*. This empirical principle indicates that firms, cities, regions, and countries are more likely to enter new activities, such as new technologies, new products, and new industries, when they already have related activities in them (Hidalgo et al., 2018; Hidalgo, 2021). For example, by analyzing the world trade data,

Hidalgo et al. (2007) calculated the density of every country's product and found that countries are more likely to diversify their exporting products toward the products with higher density. Similarly, Jun et al. (2021) expanded the density measure into three relatedness measures and found that countries are more likely to diversify their exporting product even in bilateral trade. This density measure is also used to explore the industrial diversification pattern of a region (Neffke et al., 2011). Using the Swedish data on the product portfolios in manufacturing plants, they show that the density of related industries in a region affects the probability of success in entering a new industry.

The principle of relatedness holds in regions' entering a new technology (Kogler et al., 2013). Using US patent data, Kogler et al. (2013) found that cities are more likely to enter a new technology when the city has a higher density of the new technology. They also found that cities with a higher technology density tend to exhibit a faster technological development showing their distinctive technological trajectories. The findings of Kogler et al. (2013), Rigby (2015) showed that technological relatedness determines the path of knowledge accumulation at a city level. Although Kogler et al. (2013) and Rigby (2015) analyzed the patent data, their unit of interest was not on firms but on geographical regions.

Kim et al. (2021) examined the role of technological relatedness in firms' technological diversification at a firm level using Korean firm data. In the research stream of the principal relatedness, Kim et al. (2021) find that a firm is more likely to diversify into new technology when it already has the related technologies. However, this research covered the entire technology owned by the manufacturing firm instead of focusing on a certain type of technology, such as the I4 technology. Moreover, Kim et al. (2021) didn't examine the role of government support for firms' technological diversification. In this research, we explore the effect of technological relatedness and government supports on the firm's pattern of technological diversification at a firm level focusing on the I4 technologies.

Along with the relatedness measure, we also use another measure, *complexity*, to capture the structural characteristics of technology, as well as that of a firm's technological capability. The complexity measure represents the competencies level of the economic agents or the sophistication of their economic activity, with conserving each characteristic of the activity or agent and considering their interactions as well (Hidalgo and Hausmann, 2009). In their seminal work, Hidalgo and Hausmann (2009) suggested two types of complexity: the complexity of economic agents (countries, regions, or firms) and that of activity (production of goods and services, knowledge creation, or patenting) by looking at countries' exporting products. The complexity of a country represents the country's degree of diversification in exporting products reflecting the information about the complexity of each product. The more a country engages in diverse and complex activities, the more capable the country is. Similarly, the complexity of exporting products represents the sophistication of exporting the products with preserving the complexity information of each country. The more an activity is engaged the large number of capable countries, the more sophisticated the activity is. Until now, work in complexity have proven its effects on various outcomes, including future economic growth of countries (Hidalgo and Hausmann, 2009; Hausmann et al., 2014) for various types of economic activities, such as service (Stojkoski et al., 2016), employment (Fritz and Manduca, 2021; Wohl, 2020), technology (Petraliaet al., 2017; Balland and Rigby, 2017; Balland et al., 2019), and product (Hidalgo and Hausmann,2009; Hausmann et al., 2014; Domini, 2022). In this study, we explore the factors that affect the technological diversification of firms focusing on the effect of the technological relatedness, together with the effect of the complexity of technology.

In addition, we explore the role of the government's policy support for diversification. Government support to the R&D agenda is justified based on the following argument: private sector R&D investments are suboptimal compared to the desired societal level. Since the output of firms' R&D activities shows the characteristics of public goods, the firms are likely to underproduce technological knowledge because of the appropriability issue (Stephan, 1996). Furthermore, considering the uncertainty and indivisibilities of R&D activities, firms are reluctant to do R&D that requires high cost and substantial resources (Arrow, 1962). In this aspect, the role of government in achieving the social optimal level of R&D for technological knowledge has been emphasized (Stephan, 1996) and institutions, such as patent or government subsidies, that help achieve the optimal level of knowledge creation are often regarded the prime engine of technological progress (Cohen, Nelson and Walsh, 2000; Rockett, 2010) and further, economic development of countries (North, 1990; Acemoglu et al., 2005).

The role of government becomes more critical during the early stage of technological revolutions since the uncertainty of R&D activities tends to be higher than that of other stages (Perez, 2003b) and the demand for new products from the new technologies often lags behind the speed of technological change (Dosi, 1982). Various technologies that initiated the new technological regimes have been seeded with government support. For example, technology related to packet switch funded by the Defense Advanced Research Project Agency (DARPA) resulted in the technology of Transmission Control Protocol/Internet Protocol (TCP/IP), which was the cornerstone of information and communication technologies ICT (Greenstein, 2010). Mazzucato (2015) introduces various examples of innovation and invention that were spearheaded by the state's visible hand.

Upon the emergence of a new technological paradigm during the fourth industrial revolution, the marginal effect of government support again becomes bigger with high uncertainty of R&D. For this reason, we expect a positive and significant effect of government support on firms' I4 technologies development.

## 3. Data
### *3.1. Industry 4.0 technologies*

There exist various definitions of I4 technologies (I4T), and there is no formal classification of them (Balland and Boschma, 2019; Balland et al., 2019). For example, World Bank (2020) defined the digital Industry 4.0 technologies as technologies that belonged to the following three categories based on the underlying efficiency improvement caused by the technology group: (1) informational technologies which leverage big data and analytics (e.g., cloud computing, big data analytics, and machine learning) ; (2) operational technologies which replace labor through combine data with automation (e.g., Internet of Things (IoT), 3D printing, and smart drones); and use of and (3) transactional technologies which match supply and demand such as in digital platforms and distributed ledger technologies.

Ciffolilli and Muscio (2018) classified I4T into eight categories based on expert peer reviews, and they focused on the input of the R&D process for this classification. The categories include (1) advanced manufacturing solutions, (2) additive manufacturing, (3) augmented reality, (4) simulation between interconnected machines that optimize processes, (5) horizontal and vertical integration technologies that integrate information within the value chain, (6) industrial internet and cloud that help multidirectional communication between production processes and product, (7) cyber-security that secures network operations; and (8) big data and analytics which optimize products and processes.

Balland and Boschma (2019) focused on the output of R&D activities using patent data and categorized I4T technologies into 10 different categories. Based on the Cooperative Patent Classification (CPC) code of the OECD-REGPAT database they categorized as (1) additive manufacturing; (2) artificial intelligence; (3) augmented reality; (4) autonomous robots; (5) autonomous vehicles;(6) cloud computing; (7) cybersecurity; (8) quantum computers; (9) machine tools and (10) system integration. CPC provides one of the most precise technological classifications broken down into around 250,000 categories. To identify the patents of I4 technologies, they checked the CPC code of patents and reconstructed categories indirectly by combining sub-categories. Further, they develop their own heuristics for classifying some categories that are difficult to identify and analyze the abstracts of patent data in case these categories did not allow them to identify the I4T. Considering that this research is interested in the output of firms' R&D activities, we follow the definition and classification of Balland and Boschma (2019).

*3.2 Financial information of a firm*

This study focuses on the listed Korean firms in the manufacturing sector, whose share of production in GDP is above 25% of the GDP of Korea in all Korean production. Among them, we narrowed down our research boundaries to the firms listed in three stock markets; (i) KOSPI (Korea Composite Stock Price Index), a major stock market that targets large companies, (ii) KOSDAQ (Korea Securities Dealers Automated Quotations) that targets promising small and medium-sized enterprises (SMEs) or venture companies, and (iii) KONEX (Korea New Exchange), a stock market for SMEs with lower listing thresholds. Considering that we only exclude the derivatives market from all Korean capital markets, we cover most of the firms that are listed in Korea.

For firms' financial information, we use *Kis-value* as the main dataset. This dataset from the National Information & Credit Evaluation Inc. (NICE) provides various information for external audit firms. The data include firms' general information (e.g., found year, number of workers, listed or de-listed dates) and financial information (e.g., financial statement, income statement, statement of cash flows, valuation). Key financial ratios such as debt and profit ratios can be calculated based on the *Kis-value* dataset.

*3.3 Technology(patent) information*

We use the European Patent Office (EPO) Worldwide Patent Statistical Database (PATSTATS) as our dataset for technology. The PATSTATS dataset was created by the EPO on request from the OECD and is updated twice a year (Kang and Tarasconi, 2016). In this study, we used the Spring 2021 edition to cover the years from 1984 to 2021. PATSTATS covers data from more than 90% of the world's patent authorities, including the Korean Intellectual Property Office (KIPO), United States Patent and Trademark Office (USPTO), and European countries' Patent Office, together with the European Patent Office (EPO). Using the PATSTATS database, we can access patents registered in Korea, the US, and European patent office. The data contain comprehensive information on each patent, including applicants, inventors, publications, citations, filing country, filing date, registration status, and CPC codes. In this paper, the CPC code is used as a proxy of technology.

Using both the patent and the financial data, we first match the patents from the PATSTATS dataset based on the applicant's information. However, the applicant's information is not a unique identifier in the patent data. Even for the same applicant, there could be variations in the written name due to spelling errors, abbreviations, the use of a non-unified company name, and the change in the company name. For this reason, matching a patent with its company owner is not always accurate. Accordingly, Hall et al. (2001), Thoma et al. (2007), Julius and De Rassenfosse (2014), and He et al. (2018) aimed to standardize the names of applicants, with Kim et al. (2016), Lee et al. (2019), and Kang et al. (2019) focusing on Korean firms. This study uses the OECD Harmonised Applicant Names (HAN) database, which is based on text-matching algorithms such as the one presented by Kang et al. (2019). The HAN database provides a unique firm identifier called the HAN-ID that harmonizes the names of applicants in different countries. However, there are still mismatches and errors in OECD HAN, and no changes in the company name are considered (Kang et al., 2019). This study, therefore, unifies both mismatches in the HAN ID and errors in applicants' names to obtain the name of the representative applicant. Next, this representative applicant's name is matched with the *KIS-value* data that provides financial information. As a result, we can construct a unique unbalanced panel dataset of listed Korean manufacturing companies. A total of 1,196 KIS-value firms are matched with various applicant names within PATSTATS, therewith 388,454 patent applications from 2005 to 2018.

*3.4 NTIS data*

We use data on whether firms involved in knowledge activities in I4T received direct government support. In general, government supports firm's R&D activities in two ways. One is through direct support, including grants or matching grants to targeted firms. The other is by providing indirect support through R&D tax incentives and other means, which tend to generate incremental innovation. The data from the Korean Ministry of Science and ICT (MSIT) allows us to investigate the role of the government in firms' knowledge activity in new industry 4.0 technologies. In this study, we only considered whether or not the direct government support was allocated to firms without considering other information, such as the amount of the support or the characteristics of the government-funded projects. We then checked the outcome of each government-funded project at the National Science Technology Information Service (NTIS), which is managed by the Korean MSIT. Among various proxies that show the outcome of the project (e.g., success or failure of projects, journal publication, patent, etc.), we chose the information of applied or granted patent as the outcome of each government-supported project. We gathered 696,293 applied or granted patents from the 99,482 government-funded projects between 2007 and 2018. Among them, 568,447 patents are selected.[3] Next, we checked the year of patenting and regarded it as the outcome of the government support affecting the technological development of the firm from starting year of the project. In this way, we found a total of 34,947 cases where firms develop new technologies based on the CPC code from 2007 to 2018.

## 4. Method

*4.1 Measuring technological relatedness*

We introduce a measure of relatedness and use it to estimate the proportion of related technologies already existing in the firm (Hidalgo et al., 2007; Kogler et al., 2013; Boschma et al., 2014; Gao et al., 2021; Jun et al., 2019; Kim et al., 2021). First, we connect the CPC codes of each patent and firm by building a CPC code–firm bipartite network in which the weight of the link is the number of CPC codes possessed by the firm. Every patent requires one or more CPC codes to classify its technology. We examine all the CPC codes written in one patent instead of using the representative

---

[3] When there exist multiple contributors of the outcomes of government project, each contributor claims and reports their share of contribution to the government. We regard an organization owning the patent when the report of the government-funded project claims that the share of contribution of the organization is above 50%.

CPC code. For example, if firm *i* obtains a patent applied for or granted at time *t*, whose CPC codes are A, B, and C according to the KIPO, we regard this case as the company has developed all three technologies at time *t*. In the case of a family of patents, although the patent is identical, a new CPC code (e.g., technology D) could be added following the request of an examiner when filing the patent at another patent office (e.g., USPTO). After considering a patent granted by the (for example) KIPO or USPTO as one family patent, all non-overlapping technology areas (CPC codes A–D) are synthesized and regarded as the technology of the corresponding patent.

Next, we examine the co-occurrence of the CPC codes within the patents of the same firm and then estimate proximity ($\varphi_{\alpha,\beta}$) between technologies $\alpha$ and $\beta$ following the method of Hidalgo et al. (2007). The proximity $\varphi_{\alpha,\beta,t}$ indicates the minimum value of the pairwise conditional probability that two technologies have a comparative advantage together within the same firm:

$$\varphi_{\alpha,\beta} = \min\{\Pr(RTA_\alpha|RTA_\beta), \Pr(RTA_\beta|RTA_\alpha)\} \qquad (1)$$

where RTA represents the revealed technological advantage:

$$RTA_{i,\alpha,t} = \frac{\frac{P_{i,\alpha,t}}{\sum_\alpha P_{i,\alpha,t}}}{\frac{\sum_i P_{i,\alpha,t}}{\sum_i \sum_\alpha P_{i,\alpha,t}}} \qquad (2)$$

Where $P_{i,\alpha,t}$ is the number of patents related to technology $\alpha$ possessed by firm $i$ at time $t$ (Balassa, 1965).

RTA indicates the comparative advantage of firm *i* in technology α by measuring whether it owns more technology α, as a share of its total technologies than the average firm. We state that firm *i* develops a comparative advantage in technology α at time *t* when its $RTA_{i,\alpha,t}$ has a transition from $RTA_{i,\alpha,t} < 1$ to $RTA_{i,\alpha,t} \geq 1$. Considering that previously developed technologies require some time (often more than three years) to affect the next generation of new related technologies, we examine the three years before the development of new technology. When we define the firm's development of a new technology α at time *t*, it has the technology whose RTA value is below 1 at time *t* − 1, over 1 at time *t*, and holds its value over 1 at time *t* + 1 and time *t* + 2 considering forward condition (Bahar et al., 2014).

Lastly, using the proximity among technologies, we aggregate the related technologies of a firm, which we term the *density* of the related technology of firm *i* ($\omega_{i,\alpha,t}$). It measures how a firm's existing technology portfolio is related to technology α, among all technologies. Formally, the density of the related technology for technology α of firm *i* at time *t* is given by

$$\omega_{i,\alpha,t} = \frac{\sum_\beta \varphi_{\alpha,\beta,t} U_{i,\beta,t}}{\sum_\beta \varphi_{\alpha,\beta,t}} \quad (3)$$

where $\varphi_{\alpha,\beta,t}$ is the proximity between technology $\alpha$ and $\beta$ and $U_{i,\beta,t}$ takes 1 if firm *i* has an RTA in technology $\beta$ in year *t* ($RTA_{i,\alpha,t} \geq 1$) and 0 otherwise.

*4.2 Measuring technology complexity*

Along with the relatedness measure, we also use another measure, *complexity*, to capture the structural characteristics of technology, as well as that of a firm's technological capability. The concept of complexity was introduced by Hidalgo and Hausmann (2009). In their seminal work, they focused on the economic complexity of products and countries by looking at the world trade data. Based on their observation that some countries export various kinds of products and others could export a few kinds of products only, they asked "what are the distinguishing characteristics of these two different kinds of countries and their exported products (Hidalgo and Hausmann, 2009)." To answer this, they introduced a methodology called *Method of Reflection (MOR)*. From a bipartite network, this methodology can reduce the information of one dimension (for example, producing a sophisticated product) while preserving the rest of the information of the opposite dimension (for example, products that are produced by a country whose economy is sophisticated). As a result, the *MOR provides* two symmetric information independently, about (1) actors (at country, region, city, firm, and others.) and (2) activities (industry, product, technology, occupation, and others.) which consist bipartite network.

First, *Location Complexity Index (LCI)* represents the complex level of a location by measuring how much the location has a comparative advantage in various economic activities, reflecting the information on the complexity of economic activities. According to the work of (Hidalgo and Hausmann, 2009), the countries with diversified export product portfolios are highly correlated with a high level of *LCI*. It is because countries that are more likely to export various products can produce more complicated products, which cannot easily be manufactured by many countries. Whereas the countries with few kinds of export products, their *LCI* is low as they only produce just a couple of products that are ubiquitously existed and can be manufactured by many other countries. Second, *Economic Complexity Index (ECI)* implies how frequently a certain activity is participated by various locations with preserving the location's diversification information. We can intuitively understand that economic activities are more sophisticated when only a small number of countries can develop and possess them. On the contrary, if every country can participate in and practice a certain economic activity, then that economic activity would have a low level of difficulty, in other words, a low level of complexity.

By using *MOR*, we can reduce two different, but connected information (location or economic activity from bipartite network) to the information of only one dimension by recursively calculating the average of diversification and ubiquity. The following equations are the general forms of *LCI* and *ECI*. As we have interests in technology as economic activity and firm for the location, we unify our expression of technology with subscript *t* and firms with *f*.

$$Complexity\ of\ firm: K_{f,N} = \frac{1}{K_{f,0}} \sum_t M_{f,t} K_{t,N-1} \tag{4}$$

$$Complexity\ of\ technology: K_{t,N} = TCI_{t,N} = \frac{1}{K_{t,0}} \sum_f M_{f,t} K_{f,N-1} \tag{5}$$

where $M_{f,t}$ is the matrix composed of firms and technologies with RCA above 1, which is calculated from equation (2). By using *MOR*, we can calculate the *Complexity of firm* and the *Complexity of technology*, by averaging out previous level characteristics of neighboring nodes that position at the opposite dimension iteratively. When we set two different nodes as a starting point and destination on the technology dimension, there exist numerous routes stopping by nodes of opposite dimension, a space composed of firms.

The value of *Complexity* can have different meanings based on the iteration number, $N(\geq 0)$. The iteration number *N* means how many times it iterates through nodes of different dimensions to reach the destination. The initial condition of Complexity, starting with *N* equals to 0, simply means the *degree* of a node on a network, the number of links connecting with the other nodes within the opposite dimension.

$$Diversification\ of\ firm: K_{f,0} = \sum_t M_{f,t} \tag{6}$$

$$Ubiquity\ of\ firm: K_{t,0} = \sum_f M_{f,t} \tag{7}$$

$K_{f,0}$ and $K_{t,0}$ mean technological diversification of firm (the number of technologies that are developed by each firm) and ubiquity of technology (the number of firms that develop a certain technology), respectively. As *N* increases, we average out more so that the value of *Complexity of firm* and *Complexity of technology* are converged. As it is iterated with incremented *N* until we cannot get any additional information, we stop our interaction at the 20th run following the rule of thumb of this methodology.

## 5. Results

### *5.1 Understanding technological diversification toward I4 technologies of Korean firms*

Before moving to the empirical analysis that examines the factors that affect Korean firms' knowledge activity in I4 technologies, we visualized the technology space in detail.

Figure 1(A) depicts the position of 10 different I4Ts of Korean manufacturing firms in the technology space. The 10 I4Ts are highlighted in green and related technologies are in gray. The visualization of the technology space of Korea's manufacturing firms, in general, can be found in the Appendix. As seen in the figure, all 10 I4Ts are located at the core or coherent cluster in the upper-left corner of the technology space. The size of the node is proportional to the number of patents, and among the 10 I4Ts, quantum computers, cloud computing, and cybersecurity seem to be the major patents for Korean manufacturing firms. (See Appendix A for technology space covering all the CPC codes,)

In Figures 1 (B) and (C), we explore the relationship between I4Ts. The proximity between the technologies is represented by the thickness of the edge in Figure 1 (B), and by the colors of the heatmap in Figure 1 (C). We can observe that there exists a high level of proximity between cloud computing and cybersecurity. The second-highest level of proximity can be found among augmented reality, system integration, and autonomous robots. This indicates that technologies in cloud computing and cybersecurity are likely to share the common technological capability of firms, resulting in patenting in tandem. Likewise, high proximity among technologies in augmented reality, system integration, and autonomous robots implies that they require similar technological capabilities. In addition, machine tools have a high level of average proximity with all other technologies, meaning that technologies on machine tools are widely used with other I4Ts.

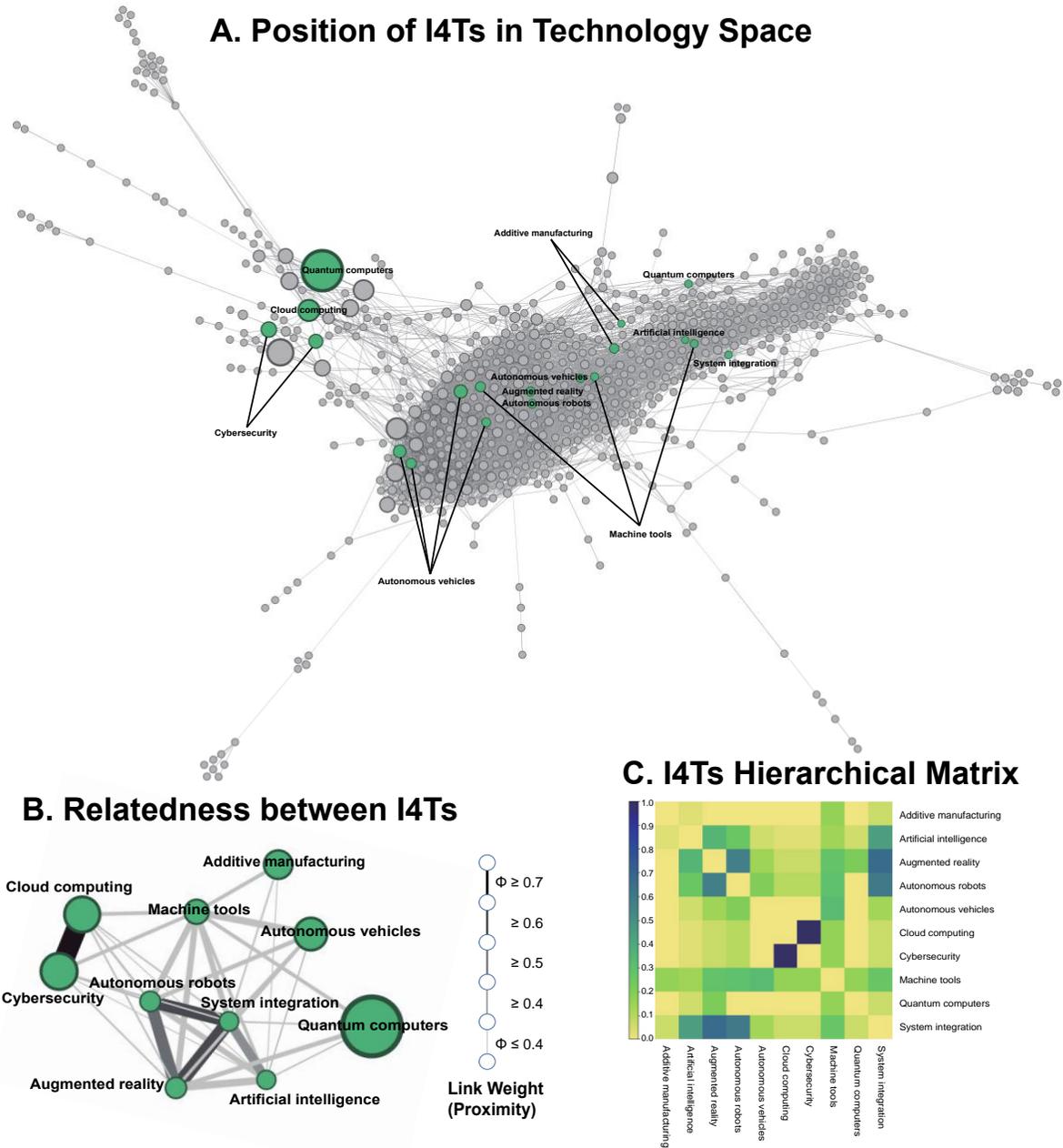

**Figure 1:** (A) Highlighting 10 different kinds of I4 technologies on the technology space of the manufacturing industry of Korea from 2010 to 2019. The green color represents CPC codes related to I4 technologies and its radius is proportional to the number of patents belonging to each CPC code. (B) A more detailed technology space composed of only 10 I4 technologies. The thickness of links means the proximity between two technologies in a pair, an alternate expression of the following hierarchical matrix. (C) Hierarchical matrix for the proximity of technology relatedness. The intense color means the high level of relatedness between two technologies in a pair.

Next, to study the sophistication of I4 technology, we aggregate patents for the recent 12-year (2007–2018) and explore their ranks of technology complexity using Equation (5). Figure 2 shows the complexity ranking of I4Ts and how they evolved over 12 years. We aggregated patents of every CPCs within each I4Ts technology and positioned it as an x-axis after log transformation. We averaged the ranking of the complexity of every CPCs within each I4 technology if multiple CPCs are assigned within one I4 technology and plotted it on the y-axis. Then, we check how the complexity ranking and the number of the patent evolved over the period into 3 phases; period1 (2007-2010), period2 (2011-2014), and period3 (2015-2018). The financial crisis happened in period1. Period2 is the recovery phase of the crisis. In period3, the seminal paper in AI was introduced (LeCun, Bengio, and Hinton, 2015) and the phenomena associated with the Forth Industrial Revolution started to be vividly observed.

A large value on the y-axis means the technology is ubiquitous with many other firms already having developed the technology. For example, from period 1 to 3, artificial intelligence became more ubiquitous, while its total number of patents is still low. Total applied patents of 9 I4Ts increase from period 1 to 3, except for quantum computers, whose total number of patents remains invariant. In the first period (2007-2010), the most commonly possessed I4T by the Korean manufacturing firms was quantum computers and the least common one was artificial Intelligence. A small number of patents in artificial intelligence technology are owned by a relatively small number of firms, while a relatively bigger amount of patents in quantum computing are developed by a larger number of firms. The most non-ubiquitous technology of period 3 was autonomous vehicles as this field was nascent with only a few firms successfully developing the technologies related to the autonomous vehicle.

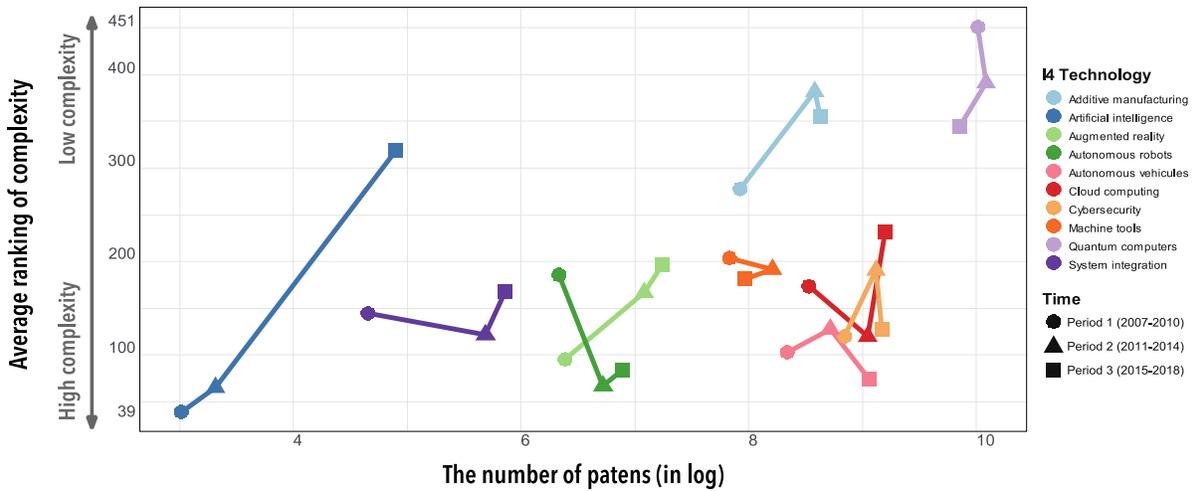

**Figure 2:** Mapping the ranking of complexity and number of patents of I4 technology. The x-axis represents the total number of patents (in log) and the y-axis represents the average ranking of the complexity of CPCs within each I4 technology. A large value on the y-axis means that the complexity of I4 technology is rather low, namely ubiquitous. The different type of I4 technologies are divided into ten colors and the shape of nodes represent the change of their coordinate value at three periods.

| Artificial intelligence | Additive manufacturing | Autonomous robots | Cybersecurity | Cloud computing |
|---|---|---|---|---|
| 1. LG Electronics Inc. | 1. LG Chem. Ltd. | 1. LG Electronics Inc. | 1. LG Electronics Inc. | 1. LG Electronics Inc. |
| 2. SK Hynix | 2. POSCO | 2. Samsung Heavy Industries | 2. Pantech | 2. SK Hynix |
| 3. Hyundai Motor Company | 3. LG Electronics Inc. | 3. Hyundai Mobis, 3. Daewoo Heavy Industries Ltd., 3. Korea Shipbuilding & Offshore Engineering, 3. SewonCorp | 3. SK Hynix | 3. LG Display |
| **Machine tools** | **Augmented reality** | **Autonomous vehicles** | **Quantum computers** | **System integration** |
| 1. Hyundai Motor Company | 1. LG Electronics Inc. | 1. Hyundai Motor Company | 1. SK Hynix | 1. LG Electronics Inc. |

| 2. Doosan Infracore | 2. Hanwha Aerospace | 2. Mando Corporation | 2. LG Display | 2. LS ELECTRIC Co., Ltd |
| 3. LG Electronics Inc. | 3. Hyundai Motor Company | 3. Hyundai Mobis | 3. Samsung SDI | 3. Hyundai Motor Company |

**Table 1:** Top three companies in terms of the number of patents in I4 technologies

Table 1 shows the name of top three companies for each I4 technologies based on the number of patents. We find well-known Korean conglomerates such as LG Electronics, SK Hynix, and Hyundai Motor Company as key players in the I4 technology space. Interestingly, Samsung Electronics, whose number of patents is one of the tops in Korea, does not have a large number of patents in I4 technologies, compared to other big firms.

We also check the age of the firm that owns the I4 technologies. From 2008 to 2014, there are 851 firms that have at least one patent in I4 technologies among all 1,113 firms. When we only look at the year 2014, there are 645 firms that have at least one patent in I4 technologies among all 908 firms. Looking at the year 2008, the average age of I4T firms is 20.7 years, while that of all firms is 21.82. For the year 2014, the average age of I4T firms is 25.45 years, while that of all firms that own any kind of patent is 26.68 years. This indicates that firms that develop the I4 technologies are slightly younger than the rest of the firms. The p-value from the t-test that asks whether the average age of firms with I4 technologies is less than that of all firms is 0.017 with 95% confidence interval, which implies that firms that develop I4 technologies are significantly younger than the rest of the firms.

Regarding the size of the firms, we measure the size as the number of employees. From 2008 to 2014, the average number of an employee of firms that owns I4 technologies is 740, while that of all firms is 630 employees. When we only look at the year 2008, the average number of employees of firms that have I4 technologies is 966, while that of all firms in the sample is 714.5. In 2014, the average number of employees of I4T firms is 1,114 while that of all firms in the sample is 899. The p-value of the t-test that asks whether the average size of firms with I4 technologies is greater than that of all firms is 0.1645 with 95% confidence interval. This result tells us that we cannot say that firms with I4 technologies are significantly larger.

*5.2 Empirical model*

To examine the factors that affect the development of new I4 technologies in a firm, we construct the following multivariate probit model.

$$U_{i,\alpha,t+2} = \beta_0 + \beta_1 \omega_{i,\alpha,t} + \beta_2 TCI_{\alpha,t} + \beta_3 Gov_{i,\alpha,t}$$
$$+ \beta_4 \boldsymbol{Firm}_{i,t} + \beta_5 \boldsymbol{Technology}_{\alpha,t} + \theta_t + \mu_i + \varepsilon_{i,\alpha,t} \quad (8)$$

$U_{i,\alpha,t+2}$ takes 1 when firm $i$ successfully enter a new technology α at time $t+2$ and 0 otherwise. Our main explanatory variable $\omega_{i,\alpha,t}$ and $TCI_{\alpha,t}$ represent the density of related technologies and the technological complexity at time $t$, respectively. $Gov_{i,\alpha,t}$ indicates whether the firm $i$ gets the government R&D support for developing the technology $\alpha$.

The second line of equation (8) consists of control variables: $\boldsymbol{Firm}_{i,t}$ includes (i) $Age_{i,t}$, the tenure of firm $i$ since its inception, $num\_employee_{i,t}$ that allows us to control for the size effect of the firm, and $Profit\_ratio_{i,t}$ and $Dept\_ratio_{i,t}$, which are the ratio of profit to sales and the ratio of total liabilities to total assets in year $t$, respectively. $Profit\_ratio_{i,t}$ and $Dept\_ratio_{i,t}$ allow us to control for the quantitative aspects of the firm (representing its capital structure that can capture its value and expected growth). Lastly, $Number\ of\ RTA_{i,t}$ represents the total number of technologies that have revealed advantage within firm $i$. This value reflects the quantitative aspect of technological capability rather than the qualitative aspect. The other vector $\boldsymbol{Technology}_{\alpha,t}$ includes the number of competitors, $num\_competitor_{\alpha,t}$ to examine firm's technological environment. This variable is the number of firms that have a comparative advantage in technology α in the industry to which the firm belongs. This variable captures two aspects: how many learning opportunities exist in the industry and how many competitors for that technology exist in the industry. We also check other factors of the firm. Finally, we add year-fixed effects ($\theta_t$) and industry-fixed effects ($\mu_i$) to control for the nationwide time trend and time-invariant characteristics of industries, respectively. $\varepsilon_{i,\alpha,t}$ is the error term.

### 5.3 Result: Technological relatedness, complexity, and diversification

We explore the factors affecting firms' technological diversification. To estimate the effect of technological relatedness and complexity described in Equation (8), we transform all variables except for the binary variables and $TCI_{\alpha,t}$ with Box-Cox transformation by year. $TCI_{\alpha,t}$ is rescaled to 0-1 and log transformed. Table 2 shows summary statistics of variables.

| Statistic | N | Mean | St. Dev. | Min | Max |
| --- | --- | --- | --- | --- | --- |
| $U_{i,\alpha,t+2}$ | 5,410,717 | 0.002 | 0.050 | 0 | 1 |
| $\omega_{i,\alpha,t}$ | 5,410,717 | 0.006 | 1.869 | −6.758 | 5.909 |
| $TCI_{\alpha,t}$ | 5,410,717 | 3.987 | 0.527 | 2.010 | 4.615 |
| $Gov_{i,\alpha,t}$ | 5,410,717 | 0.004 | 0.062 | 0 | 1 |
| $Age_{i,t}$ | 5,410,717 | −0.00005 | 1.280 | −4.668 | 4.136 |
| $num\_Employee_{i,t}$ | 5,410,717 | 0.0002 | 0.626 | −4.142 | 2.754 |
| $num\_Competitor_{\alpha,t}$ | 5,410,717 | 0.009 | 2.208 | −5.019 | 5.376 |
| $Num\_RTA_{i,t}$ | 5,410,717 | 0.00000 | 0.804 | −1.553 | 2.176 |
| $Profit\_ratio_{i,t}$ | 5,410,717 | 0.196 | 3.044 | −47.861 | 43.226 |
| $Debt\_ratio_{i,t}$ | 5,410,717 | −0.00000 | 0.005 | −0.013 | 0.038 |

**Table 2:** Summary statistics. All variables are transformed by Box-Cox transformation by year and are pooled.

Table 3 is the correlation table for all the variables. We can see that $\omega_{i,\alpha,t}$ is heterogeneous among firms and is highly correlated with the number of other kinds of firms' existing technology. Interestingly, $\omega_{i,\alpha,t}$ is not highly correlated with a firm's basic characteristics such as firm's age($Age_{i,t}$), size($num\_employee_{i,t}$) or financial structure($Profit\_ratio_{i,t}$ and $Dept\_ratio_{i,t}$), but highly correlated with the number of technologies that has revealed technological advantage ($Num\_RTA_{i,t}$). To avoid the multicollinearity problem among independent variables, the variance inflation factor (VIF) was measured, and it was confirmed that the value was not very high at 1.219. We also tested VIF for $Age_{i,t}$ and $num\_employee_{i,t}$, and the result was 1.055, which implies that we can simultaneously consider have all variables in our regression model.

| | $U_{i,a,t+2}$ | $\omega_{i,a,t}$ | $TCI_{a,t}$ | $Gov_{i,a,t}$ | $Age_{i,t}$ | $num\_Competitor_{a,t}$ | $Num\_RTA_{i,t}$ | $num\_Employee_{i,t}$ | $Profit\_ratio_{i,t}$ | $Debt\_ratio_{i,t}$ |
|---|---|---|---|---|---|---|---|---|---|---|
| $U_{i,a,t+2}$ | 1 | | | | | | | | | |
| $\omega_{i,a,t}$ | 0.061 | 1 | | | | | | | | |
| $TCI_{a,t}$ | 0.005 | 0.001 | 1 | | | | | | | |
| $Gov_{i,a,t}$ | 0.022 | 0.090 | -0.011 | 1 | | | | | | |
| $Age_{i,t}$ | 0.021 | 0.191 | -0.00004 | 0.028 | 1 | | | | | |
| $num\_Competitor_{a,t}$ | 0.032 | 0.232 | -0.009 | 0.017 | 0.00000 | 1 | | | | |
| $Num\_RTA_{i,t}$ | 0.051 | 0.870 | 0.00001 | 0.081 | 0.206 | 0.00000 | 1 | | | |
| $num\_Employee_{i,t}$ | 0.035 | 0.343 | 0.0004 | 0.055 | 0.462 | 0.00000 | 0.352 | 1 | | |
| $Profit\_ratio_{i,t}$ | -0.002 | -0.022 | 0.071 | -0.001 | -0.075 | -0.0001 | -0.005 | -0.050 | 1 | |
| $Debt\_ratio_{i,t}$ | 0.004 | 0.031 | -0.0004 | 0.003 | -0.044 | 0.00000 | 0.024 | 0.003 | -0.130 | 1 |

**Table 3:** Correlation table

|  | Dependent variable: $U_{i,\alpha,t+2}$ | | | | |
|---|---|---|---|---|---|
|  | (1) | (2) | (3) | (4) | (5) |
| $\omega_{i,\alpha,t}$ | 0.001518*** | 0.001479*** |  | 0.001159*** | 0.001035*** |
|  | (0.000010) | (0.000010) |  | (0.000026) | (0.000028) |
| $TCI_{\alpha,t}$ |  | 0.000559*** |  | 0.000526*** | −0.000093 |
|  |  | (0.000036) |  | (0.000041) | (0.000188) |
| $Gov_{i,\alpha,t}$ |  | 0.013267*** |  | 0.012644*** | 0.011948*** |
|  |  | (0.000313) |  | (0.000343) | (0.000344) |
| $Age_{i,t}$ |  |  | 0.000160*** | 0.000172*** | 0.000208*** |
|  |  |  | (0.000018) | (0.000019) | (0.000023) |
| $num\_Employee_{i,t}$ |  |  | 0.001414*** | 0.001222*** | 0.001028*** |
|  |  |  | (0.000040) | (0.000040) | (0.000046) |
| $num\_Competitor_{\alpha,t}$ |  |  | 0.000729*** | 0.000497*** | 0.000520*** |
|  |  |  | (0.000009) | (0.000011) | (0.000011) |
| $Num\_RTA_{i,t}$ |  |  | 0.002734*** | 0.000360*** | 0.000490*** |
|  |  |  | (0.000028) | (0.000060) | (0.000063) |
| $Profit\_ratio_{i,t}$ |  |  | −0.000000 | 0.000003 | 0.000003 |
|  |  |  | (0.000007) | (0.000007) | (0.000007) |
| $Debt\_ratio_{i,t}$ |  |  | 0.02615*** | 0.022604*** | 0.011956** |
|  |  |  | (0.00393) | (0.003929) | (0.004176) |
| Industry fixed effect | No | No | No | No | Yes |
| Time fixed effect | No | No | No | No | Yes |
| Constant | 0.002321*** | 0.000052 | 0.002479*** | 0.000329** |  |
|  | (0.000019) | (0.000146) | (0.000021) | (0.000163) |  |
| Observations | 6,299,159 | 6,299,159 | 5,410,717 | 5,410,717 | 5,410,717 |
| Adjusted $R^2$ | 0.003 | 0.004 | 0.004 | 0.004 | 0.005 |
| Residual Std. Error | 0.048 | 0.048 | 0.050 | 0.049 | 0.049 |

Note: *p<0.1; **p<0.05; ***p<0.01

**Table 4:** Results covering all technologies, firms' technological diversification, 2005–2016.

Table 4 shows the results with the entire sample covering all technologies. In columns (1) and (2), only the main variables of interests are considered. The result shows that $\omega_{i,\alpha,t}$, $TCI_{\alpha,t}$, $Gov_{i,\alpha,t}$ had a positive and significant effect on firm's technological diversification. This implies that firms are more likely to develop technologies that are related to the technologies they already have. This tendency gets stronger when technology is more complex.

As shown in columns (4) and (5), the positive and significant effects hold even with control variables. Column (4) tells that the increase of technological relatedness $\omega_{i,\alpha,t}$ by 1 unit results in 0.11% increase in the odds of the developing technology $\alpha$ with other variables remain fixed. Likewise, an increase in 1 unit of technological complexity and the government support $Gov_{i,\alpha,t}$ enhance the odds of technological development by 0.05% and 1.272%, respectively.

Column (3) shows the result considering the effect of control variables only. The result shows that older and larger firms are more likely to develop new technologies. This might be because older firms have already accumulated their technological capabilities and larger firms can utilize a larger amount of R&D resources, such as R&D personnel. The probability of knowledge spillover is high when a firm has various technology pools. Lastly, firms with a higher debt ratio are more likely to diversify their technologies. The profit ratio does not make a significant effect on technological diversification.

To investigate when the technological complexity can make a significant effect on the diversification, we select the two samples based on the level of technological relatedness ($\omega_{i,\alpha,t}$). One group consists of firms with top 10% relatedness, and the other is the bottom 10% of relatedness. When we compare column (4) with column (5), the significant and positive effect of technological complexity, $TCI_{\alpha,t}$, disappears with industry and time fixed effects.

Table 5 shows the effects of technological complexity ($TCI_{\alpha,t}$) on technological diversification over different relatedness. The first column (1) of table 5 indicates that technological complexity plays more role in developing a new diversified technologies when firms already have a high level of technological relatedness. This implies that for firms to develop more complex and non-ubiquitous technologies, they need to accumulate related technologies first. Balland et al. (2019) explain it as a 'diversification dilemma' meaning that a firm cannot develop more difficult technology although it is more attractive.

Interestingly, the effect of government support disappears for the firms that have already accumulated related technologies, while the effect is significant for firms with low technological relatedness. This result implies that government support is more crucial for firms that have not yet accumulated enough technological capabilities. The positive and significant effects of government support in columns (2), (4), and (6) imply that government, with limited budget resources, should focus their support on firms that have low level of technological relatedness so that they could jump into unrelated technology spaces. Therewith more complex technologies and firms need to exploit the government R&D support to diversify their technology toward unrelated, less ubiquitous, and more complex technologies.

|  | (1) High relatedness | (2) Low relatedness | (3) High relatedness | (4) Low relatedness | (5) High relatedness | (6) Low relatedness |
|---|---|---|---|---|---|---|
|  | \multicolumn{6}{c}{Dependent variable: $U_{i,\alpha,t+2}$} | | | | | |
| $TCI_{\alpha,t}$ | 0.001243*** (0.000266) | 0.000061* (0.000034) | 0.003377*** (0.000319) | 0.000037 (0.000040) | 0.004252*** (0.001294) | −0.000318** (0.000136) |
| $Gov_{i,\alpha,t}$ | 0.004064*** (0.000941) | 0.013657*** (0.000822) | 0.000004 (0.001039) | 0.010705*** (0.000886) | −0.000609 (0.001042) | 0.010679*** (0.000886) |
| $Age_{i,t}$ |  |  | 0.001398*** (0.000160) | −0.000002 (0.000016) | 0.001874*** (0.000264) | 0.000002 (0.000021) |
| $num\_Employee_{i,t}$ |  |  | 0.001416*** (0.000263) | 0.000190*** (0.000041) | 0.000100 (0.000447) | 0.000145*** (0.000046) |
| $num\_Competitor_{\alpha,t}$ |  |  | 0.002430*** (0.000071) | 0.000107*** (0.000011) | 0.002528*** (0.000071) | 0.000107*** (0.000011) |
| $Num\_RTA_{i,t}$ |  |  | 0.010300*** (0.000694) | 0.000050 (0.000050) | 0.009348*** (0.000971) | 0.000006 (0.000058) |
| $Profit\_ratio_{i,t}$ |  |  | −0.000212*** (0.000057) | 0.000007 (0.000005) | −0.000183*** (0.000063) | 0.000009 (0.000006) |
| $Debt\_ratio_{i,t}$ |  |  | −0.087554*** (0.030818) | −0.002306 (0.003389) | −0.179470*** (0.038829) | −0.001508 (0.003744) |
| Industry fixed effect | No | No | No | No | Yes | Yes |
| Time fixed effect | No | No | No | No | Yes | Yes |
| Constant | 0.007021*** (0.001084) | −0.000058 (0.000138) | −0.018360*** (0.001832) | 0.000294 (0.000188) |  |  |
| Observations | 629,916 | 629,915 | 541,072 | 541,072 | 541,072 | 541,072 |
| Adjusted $R^2$ | 0.000059 | 0.000440 | 0.002975 | 0.000520 | 0.005262 | 0.000749 |
| Residual Std. Error | 0.109506 | 0.013969 | 0.112709 | 0.014637 | 0.112579 | 0.014635 |

Note:  *p<0.1; **p<0.05; ***p<0.01

**Table 5:** The effects of technological complexity on the technological diversification of firms over the different technological relatedness.

### 5.4 Result: Factors that affect firms' technological diversification toward I4 technologies

To examine the factors that determine the technological diversification of firms toward I4 technologies, we selected the 10 I4 technologies only and examined the effects of technological relatedness and other variables on technological diversification. We did not see the effect of technological complexity ($TCI_{\alpha,t}$) here since we already select a sample consisting of only I4 technologies, whose complexity is similar.

| | (1) | (2) | (3) Large technology stock (4th quartile) | (4) Medium high tech. stock (3rd quartile) | (5) Medium low tech. stock (2nd quartile) | (6) Small technology stock (1st quartile) | (7) High density (4th quartile) | (8) Low density (1st quartile) |
|---|---|---|---|---|---|---|---|---|
| $\omega_{i,t}$ | 0.002103*** (0.000382) | 0.007726*** (0.001608) | 0.004188 (0.003924) | 0.009228*** (0.003184) | 0.005570** (0.002701) | 0.001408 (0.001225) | | |
| $Gov_{i,t}$ | 0.005079*** (0.001597) | 0.006425*** (0.001765) | 0.006622* (0.003812) | 0.012449*** (0.004225) | −0.000484 (0.003173) | 0.006034** (0.003021) | 0.005579 (0.004055) | 0.005067** (0.002117) |
| $Age_{i,t}$ | | −0.000015 (0.000630) | −0.001853 (0.002487) | 0.002467 (0.001759) | −0.000860 (0.001125) | −0.000182 (0.000888) | 0.001016 (0.002493) | 0.000498 (0.000640) |
| num.Employee$_{i,t}$ num.Employee | | −0.001069 (0.001772) | −0.001872 (0.006286) | −0.010433* (0.005391) | 0.001448 (0.003333) | 0.001395 (0.002459) | −0.006180 (0.006742) | 0.000481 (0.001777) |
| num.Competitor$_{i,t}$ | | 0.000856** (0.000366) | −0.000807 (0.000903) | 0.001724** (0.000833) | 0.001202* (0.000627) | 0.002176*** (0.000509) | −0.000739 (0.000968) | 0.001499*** (0.000381) |
| $Num\_RTA_{i,t}$ | | −0.006490*** (0.001772) | | | | | −0.005868 (0.005271) | −0.001690 (0.001065) |
| $Profit\_ratio_{i,t}$ | | −0.000066 (0.000245) | −0.000817 (0.000726) | −0.000149 (0.000607) | 0.000039 (0.000375) | 0.000229 (0.000337) | −0.000818 (0.000810) | 0.000256 (0.000239) |
| $Debt\_ratio_{i,t}$ | | −0.196468 (0.121861) | −0.422719 (0.446120) | −0.336622 (0.292333) | 0.212875 (0.212587) | −0.128772 (0.168643) | −0.870960* (0.499131) | −0.140125 (0.120451) |
| Industry fixed effect | Yes | Yes | Yes | Yes | Yes | Yes | Yes | Yes |
| Time fixed effect | Yes | Yes | Yes | Yes | Yes | Yes | Yes | Yes |
| Observations | 21,744 | 19,133 | 4,781 | 4,782 | 4,784 | 4,786 | 4,781 | 4,786 |
| Adjusted $R^2$ | 0.000743 | 0.002909 | 0.006557 | 0.016856 | −0.003055 | 0.001871 | 0.005628 | 0.005067 |
| Residual Std. Error | 0.077942 | 0.079807 | 0.100397 | 0.086889 | 0.064628 | 0.059443 | 0.110101 | 0.043218 |

Note: *p<0.1; **p<0.05; ***p<0.01

Dependent variable: $U_{i,n,t+2}$

**Table 6:** The effect of relatedness and government support on firms' technological diversification toward I4 technologies

Table 6 shows the results. Column (1) describes the effects of the relatedness and government support, while column (2) shows the results with all the control variables. We control for the industry- and time-fixed effects for all columns. The previous results that are shown in tables 4 and 5, hold for the I4 technologies as well. Firms are likely to enter a new I4 technology when they have more related technology base, more government supports, and more competitors with the same technologies. Firms' age does not matter for the technological diversification toward the I4 technologies. However, the total number of technologies of the firm, $Num\_RTA_{i,t}$, negatively affect the diversification, implying that firms with the smaller stock of technologies perform better at the diversification toward the I4 technologies.

With intrigued by the results in column (2), in Column (3)-(6), we split the sample over the level of technology stock, which is measured by the total number of competitive technologies of the firm and see the results focusing on the density of the technologies and government support. Regarding the effect of the density, only firms with medium-high and medium-low technology stock exhibit the significant effect of the density on their I4T diversification. Since the firms with small technology stock, by definition, lack the related technology for their diversification, the effect of related technology seems not to be significant. In the case of the firms with large technology stock, since they already occupied a vast territory in the technology space, their marginal expansion of technology may be difficult so that the effect of the density turns out to be not significant.

Regarding the government support, interestingly, the effect of government support is the strongest for firms with medium-high technology stock. Next, its impact seems to be meaningful for firms with small and large technology stock. For firms with a smaller stock of technologies, the related technologies seem to compensate for the capabilities that are associated with the number of technologies, while the efficiency of government support seems to be the greatest for firms with medium-high technology stock.

To check the effectiveness of government support, we split the sample over the different levels of density of I4 technologies in Columns (7) and (8). The effects of the density of I4 technologies and government support are only significant for firms with low density. This result indicates that the support from the government can help the firms that do not have the related technology yet to jump into the I4 technologies.

To examine the different effects of related technologies and government support over different sizes of firms, we split the sample over the number of employees, which is the proxy of firms' size in Table 7. First, the positive effect of the density is the strongest for the big firms and strong for the smaller firms, while the significant effect disappears for the firms in the $2^{nd}$ quartile of size. Second, government support is not working well for big firms. This indicates that government direct support increases the probability of success for small firms to enter the new I4 technologies.

|  | Dependent variable: $U_{i,\alpha,t+2}$ | | | |
|---|---|---|---|---|
|  | (1) 1st quartile | (2) 2nd quartile | (3) 3rd quartile | (4) 4th quartile |
| $\omega_{i,\alpha,t}$ | 0.006679*** | 0.005731 | 0.009482** | 0.009697* |
|  | (0.002508) | (0.003596) | (0.004104) | (0.005841) |
| $Gov_{i,\alpha,t}$ | 0.000585 | 0.007294** | 0.006858* | 0.007872** |
|  | (0.003163) | (0.003670) | (0.003995) | (0.003687) |
| $Age_{i,t}$ | −0.001245 | 0.001242 | 0.003294** | −0.003582* |
|  | (0.001103) | (0.001445) | (0.001354) | (0.002003) |
| $num\_Competitor_{\alpha,t}$ | 0.003644*** | 0.004190*** | 0.001767* | 0.000422 |
|  | (0.000814) | (0.001010) | (0.001049) | (0.001083) |
| $Num\_RTA_{i,t}$ | −0.006183** | −0.004859 | −0.007588* | −0.006669 |
|  | (0.002772) | (0.003892) | (0.004438) | (0.006599) |
| $Profit\_ratio_{i,t}$ | 0.000025 | 0.000131 | −0.000552 | −0.000195 |
|  | (0.000363) | (0.000468) | (0.000573) | (0.000673) |
| $Debt\_ratio_{i,t}$ | −0.109596 | 0.018814 | −0.148580 | −0.476133 |
|  | (0.172473) | (0.270223) | (0.329247) | (0.384645) |
| Artificial_intelligence | −0.000810 | 0.005501 | −0.006999 | −0.009368 |
|  | (0.014639) | (0.019118) | (0.018784) | (0.011665) |
| Augmented_reality | 0.005005 | −0.001795 | −0.006760 | −0.013352** |
|  | (0.005264) | (0.006483) | (0.008495) | (0.006787) |
| Autonomous_robots | −0.006266 | 0.005213 | 0.004788 | −0.003400 |
|  | (0.006405) | (0.006802) | (0.007578) | (0.006978) |
| Autonomous_vehicles | −0.003680 | −0.005112 | −0.005091 | −0.005424 |
|  | (0.004166) | (0.005031) | (0.004625) | (0.004660) |
| Cloud_computing | 0.009922** | 0.008139* | 0.007978 | 0.006737 |
|  | (0.003911) | (0.004799) | (0.005020) | (0.006063) |
| Cybersecurity | 0.013137*** | 0.011604** | 0.014093** | 0.005314 |
|  | (0.004403) | (0.005166) | (0.006024) | (0.005955) |
| Machine_tools | 0.005982* | 0.013899*** | 0.011601*** | −0.000554 |
|  | (0.003618) | (0.003807) | (0.003791) | (0.004520) |
| Quantum_computers | 0.016281*** | 0.014532*** | 0.009181** | 0.002000 |
|  | (0.003705) | (0.004307) | (0.004411) | (0.005081) |
| System_integration | −0.007000 | −0.005197 | −0.007803 | −0.012733 |
|  | (0.008255) | (0.010411) | (0.011176) | (0.008686) |
| Observations | 4,786 | 4,784 | 4,782 | 4,781 |
| Adjusted $R^2$ | 0.015651 | −0.000461 | −0.003545 | 0.009832 |
| Residual Std. Error | 0.065583 | 0.077648 | 0.082940 | 0.090645 |

Note: *p<0.1; **p<0.05; ***p<0.01

**Table 7:** Sample split over firms' size. The different effects of relatedness and government support on firms' technological diversification toward I4 technologies over the firms' size that is measured by the number of employees

## 6. Conclusion

This research analyzed the factors that affected the firms' technological diversification toward the I4 technologies using data on patents, financial information, and government-funded projects of Korean manufacturing firms. In terms of the absolute number of patents owned by Korean manufacturing firms, dominant technologies among the I4 technologies were quantum computers, cloud computing, and cybersecurity. As we checked the relationship between the I4 technologies by constructing a technology space, we also found that cloud computing and cybersecurity technologies have the closest relationship, implying that those two technologies share capabilities in common. Moreover, augmented reality, system integration, and autonomous robot showed high proximity to each other. The technology of the autonomous vehicle turned out to be relatively non-ubiquitous, although relatively a large number of patents was filed and obtained for this technology. This implies that only a few companies can develop technologies related to autonomous vehicles, and they occupy a monopolistic status in this technology space. When we looked at the name of the firms that published patents in I4 technologies, well-known big firms mostly developed those technologies in Korea.

Firms that already have the related technologies are more likely to enter new I4Ts. This result means that when a firm aims to enter one of each technology, having other related technologies can increase the success rate in entering the technology. The result suggests that firms need to accumulate related technologies to develop more complex and non-ubiquitous technologies.

Government direct support increases the probability of success for firms to enter the technology space. The government support turns out to be more significant for firms that possess low technological relatedness and firms that possess a medium-high and small level of technology stock when entering the I4 technologies. The result shows that firms that receive government R&D support can diversify their technology toward unrelated, less ubiquitous, and more complex technologies.

Interestingly, the role of government support diminishes as firms accumulate more related technologies. This implies that firms that already have accumulated the related technologies do not depend on government support for entering a new technology. It seems that the role of government support is more of an instigator or a nudge for firms that are relatively less competitive with low accumulated technologies. Government can maximize its resources by supporting firms with low relatedness and small technology stock so that they could jump into unrelated, and more complex technologies. The role of government support in firms' diversifying their technologies toward I4 technology can be to nudge firms so that they can be better equipped to climb the ladder of technology by themselves rather than to hand over the ladder itself to the firms.

The fact that big firms are more likely to enter those technologies indicates that entering I4 technologies is not an easy task for smaller firms with limited R&D resources. Therefore, policies should be designed and implemented considering the context of each country's firm and market. At the same time, government support needs to be oriented towards supporting the firm's technology accumulation in general rather than focusing on leapfrogging into the I4 technologies. This is because the positive effect of government support disappears when the firm size becomes large.

We expect that this research is helpful for policymakers, who aim to diversify their technology and gain competitiveness in industry 4.0 technologies. One caveat is that this paper is solely based on the Korean manufacturing sector. Big manufacturing firms are the main players that drive the technological transition toward I4 technologies in Korea. This should be understood in the context of Korea's history of economic and industrial development.

**Acknowledgment**
This paper is based on research funded by the World Bank Seoul Center for Finance and Innovation and by the 2021 Research Fund (1.210036.01) of UNIST.

**Appendix A. Building a technology space of Korea for 2010–2019**

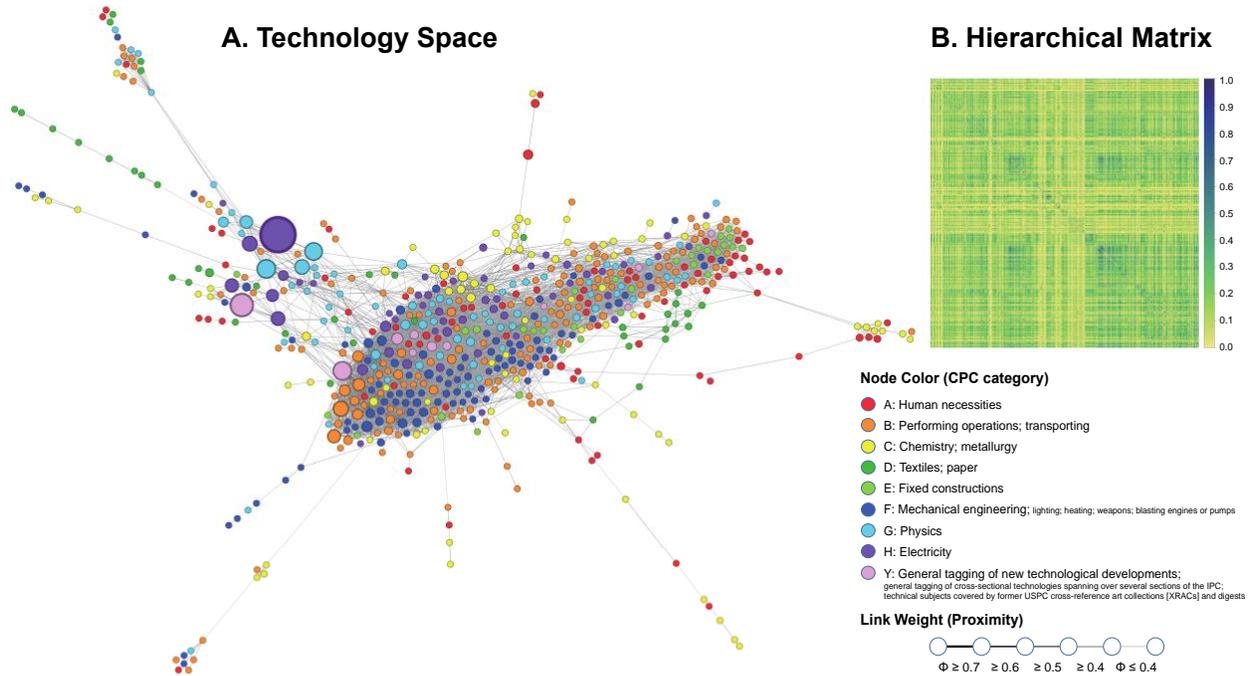

**Figure** A1: (A) Technology space of manufacturing industry of Korea from 2010-2019. The color of the node is classified by 1-digit CPC codes and its radius is proportional to the number of patents belonging to each CPC codes. (B) Hierarchical matrix for the proximity of technology relatedness *Source: Kim et al. (2021)*

We draw Figure A1, which is the technology space of Korea based on the patent data from 2010 to 2019. Each node represents a technological category expressed with CPC code, while the links represent the proximity between technologies. The radius of nodes can be viewed as broadly indicative of total number of patents in that CPC codes. The technology space of Korea exhibits a core/periphery structure such that the subcategories of physics and mechanical engineering, lighting, heating, and weapons are likely to be located in the center, while those of human necessities, textiles, and paper are at the periphery. Simultaneously, we can observe a coherent cluster in the upper left-hand corner, which consists of the subcategories of electricity, physics and mechanical engineering, lighting, heating, and weapons. This cluster is from Samsung, which owns 31% of all Korean patents and specializes as an electronics and chip maker.